# Ultrasensitive Nanoplastics Detection Leveraging Shrinking Surface Plasmonic Bubble

Yang Liu[1,†], Renzheng Zhang[1,†], Amartya Mandal[1], Eungkyu Lee[4], Seunghyun Moon[1*], and Tengfei Luo[1,2,3*]

[1]Department of Aerospace and Mechanical Engineering, University of Notre Dame, Notre Dame, IN, USA

[2]Department of Chemical and Biomolecular Engineering, University of Notre Dame, Notre Dame, IN, USA

[3]Center for Sustainable Energy of Notre Dame (ND Energy), University of Notre Dame, Notre Dame, IN, USA

[4]Department of Electronic Engineering, Kyung Hee University, Yongin-si, Republic of Korea

*E-mail: S. Moon <smoon7@nd.edu> and T. Luo <tluo@nd.edu>
†These authors contribute equally to this work.

**Abstract:**
Nanoplastics pose serious environmental and health risks due to their widespread presence in aquatic systems. Detecting trace amounts of nanoplastics is a challenging task, which currently requires sophisticated equipment and tedious sample preparation (e.g., ultrafiltration). In this work, we demonstrate an ultra-sensitive Shrinking Surface Bubble Deposition (SSBD) technique for nanoplastics detection. SSBD leverages plasmonic photothermal effects to generate a surface bubble and the resulting Marangoni flow to concentrate sparsely suspended nanoplastics onto the bubble surface. The collected nanoplastic particles are subsequently deposited on the substrate after the bubble shrinks and vanishes. To quantify the detection limit of SSBD for nanoplastics in water, core–shell gold plasmonic nanoparticles are mixed with the aqueous sample to enable photothermal bubble generation, while also supporting surface-enhanced Raman spectroscopy (SERS) for signal enhancement. Results show that the limits of detection are 10 ng/mL, $10^{-1}$ ng/mL and $10^{-3}$ ng/mL for polystyrene (PS) particles with diameters of 500 nm, 200 nm and 30 nm, respectively. We further used SSBD to detect plastics particles in real drinking water (e.g., bottled and fountain water) and found polyamides (PA) and polypropylene (PP) micro/nanoplastics, demonstrating the potential of the SSBD-SERS technique as a versatile and sensitive platform for detecting trace-level nanoplastic contamination and assessing human exposure risk.

The major sources of plastic particles include industrial waste and fragments from daily-use products, which undergo physical and chemical degradation [1] [2] and enter aquatic environments through soil, water, and air pathways [3]. Over time, larger debris breaks down into micro- and even nanoparticles (NPs) (< 1 μm) [4], increasing their number density in rivers, lakes, coastal zones, and even open oceans [5] [6]. Due to their small size, high surface area, and chemical composition, nanoplastics pose significant ecological and health risks [7]. They can accumulate in tissues, cross biological barriers, and adsorb pollutants, leading to oxidative stress, hormonal disruption, and impaired reproduction in fish [8] [9] [10] [11] [12]. Nanoplastics can also penetrate the blood–brain barrier, causing neuroinflammation and behavioral changes in animals, and elevated levels have been detected in human brain tissue, suggesting potential long-term health impacts [13] [14] [15] [16]. Consequently, their widespread occurrence has raised growing concerns about environmental and human health.

Techniques for detecting nanoplastics in water have gained increasing attention [17] [18]. Conventionally, nanoplastics in large-volume environmental samples must first be isolated and concentrated using various methods such as ultrafiltration, evaporation, solvent extraction, or membrane-based pre-concentration. Only after these pre-treatment steps can existing techniques, mainly pyrolysis–gas chromatography–mass spectrometry (Py-GC/MS) [19] [20] [21], be used to chemically identify them. However, Py-GC/MS is a destructive method hence cannot provide particle size or morphology information, which are essential for understanding their environmental behavior and toxicity. Raman spectroscopy provides molecular-level information and allows non-destructive analysis of nanoplastics in complex matrices such as sediments, tissues, and living organisms [22] [23] [24] [25]. Surface-Enhanced Raman Spectroscopy (SERS) further increase the sensitivity to allow trace-level detection by exploiting localized surface plasmon resonance in metallic nanostructures (e.g., gold, Au; silver, Ag), thereby amplifying Raman signals by orders of magnitude [26] [27]. This enhancement enables the detection of nanoplastics in water, biological fluids, food, and other complex matrices at trace concentrations, while also providing molecular-level insights into polymer composition and degradation through spectral shifts or band broadening [28] [29].

Recent advancements in SERS technology have greatly expanded its applicability in nanoplastics detection. Lê et al. designed anisotropic nanostar dimer–embedded nanopore SERS substrates for the detection of submicron polystyrene (PS) particles in water, achieving a detection limit of 50 mg/L for 400 nm particles within minutes [30]. Xu et al. employed a commercial Klarite-based SERS substrate composed of inverted pyramidal Au cavities, achieving enhancement factors of up to two orders of magnitude for the Raman signal and enabling single-particle detection of PS and polymethyl methacrylate (PMMA) down to 360 nm at a concentration of 26.25 μg/mL [31]. They further demonstrated the detection and identification of nanoplastics as small as 450 nm extracted from environmental samples. Ruan et al. developed a sol-based SERS method using Ag NPs as an aggregation medium to detect nanoplastics in aqueous environments, achieving a detection limit of 5 mg/L and enabling the detection of plastic particles as small as 20 nm within 2 minutes [32]. Recently, Shi et al. developed a method combining optical manipulation with SERS (OM–SERS) using 20 and 80 μm gold NP stacks to simultaneously manipulate, enrich, and detect nanoplastics through the photothermal effect. This approach achieved high enrichment recoveries (89.3–94.3%) and low detection limits (150 ng/L) for PS nanoplastics [33]. In addition, Qian et al. developed a stimulated Raman scattering (SRS) imaging platform with automated plastic identification, enabling single-particle micro- and nanoplastic analysis. They further demonstrated that narrowband SRS enhances sensitivity, allowing rapid single nanoplastic detection below 100 nm [25]. In our previous work, Moon et al. developed a Shrinking Surface Bubble Deposition (SSBD) method for nanoplastics enrichment and detection in environmental water samples and identified diverse nanoplastic chemistry and morphologies from samples collected all around the world oceans [34]. It only needed ~1 ml of water sample and does not need extensive pretreatments. However, the limit of detection (LOD) of SSBD is not known, obscuring its application perspective.

In this work, we quantify the LOD of the SSBD technique using PS nanoplastics spiked in water as benchmarking samples before demonstrating it on plastics detection in real drinking water samples. Our results show that, in the commonly used metric, gram of plastics per liter of water sample (g/L), the LOD of SSBD mainly depends on the particle size. For 500 nm PS particles, SSBD detect at concentrations as low as 10 ng/mL, while 30 nm PS particles are detectable at concentrations down to $10^{-3}$ ng/mL, which is ~150 times lower than the most sensitive nanoplastics detection technique reported to date [33]. When used to detect plastics particles in real drinking water (e.g., bottled and fountain water), we found polyamides (PA) and polypropylene (PP) micro/nanoplastics. This study demonstrates the potential of the SSBD technique as a platform for detecting trace-level nanoplastic contamination, which is critical for assessing environmental and human exposure risk.

**SSBD working principle**
In the SSBD process, core–shell gold plasmonic nanoparticles ($SiO_2$@Au NPs) are mixed with the aqueous sample to enable photothermal bubble generation upon laser illumination (Fig. 1a and Supplementary Fig. 1), while also supporting SERS for signal enhancement (Fig. 1b). SSBD leverages the Marangoni flow around the photothermal bubble on the substrate surface to concentrate suspended nanoplastics at the bubble surface together with the plasmonic nanoparticles [34] [35]. Specifically, an 800 nm pulsed laser irradiates a suspension of $SiO_2$@Au NPs (120 nm $SiO_2$ core diameter, 15 nm Au shell thickness) contained in a transparent PMMA cuvette (Fig. 1a). Optical pressure drives the $SiO_2$@Au NPs toward the glass surface, where they act as surface heaters and nucleation sites for surface bubbles formation (Fig. 1c) [36] [37] [38]. The laser wavelength (800 nm) closely matches the surface plasmon resonance peak of the $SiO_2$@Au NPs (~760 nm), ensuring efficient plasmonic heating (Supplementary Fig. 2). As laser heating continues, bubbles grow. The localized heating creates a temperature gradient, with the region above the bubble hotter than the bottom (Figs. 1d,e), thereby resulting in Marangoni flow. The flow brings NPs toward the bubble surface, where they are captured by the balance between inward surface tension forces and outward pressure at the bubble–liquid interface [39] [40]. Once trapped, $SiO_2$@Au and PS NPs are transported by Marangoni flow to the three-phase contact line (TPCL). When the bubble diameter reaches a critical size (~40 μm), the laser is switched off. Cooling causes bubble shrinkage, and upon complete disappearance, a densely packed NPs deposition remains on the surface, unless the bubble detaches from the glass (Fig. 1c). The SSBD process completes within 10 min. The resulting deposits can be subsequently characterized using techniques such as scanning electron microscopy (SEM) and Raman spectroscopy to identify the morphology and chemistry of the deposited particles [41] [42]. The co-deposition of the plasmonic NPs and the plastic particles enables the SERS effect, which can lead to the detection of trace-level nanoplastics.

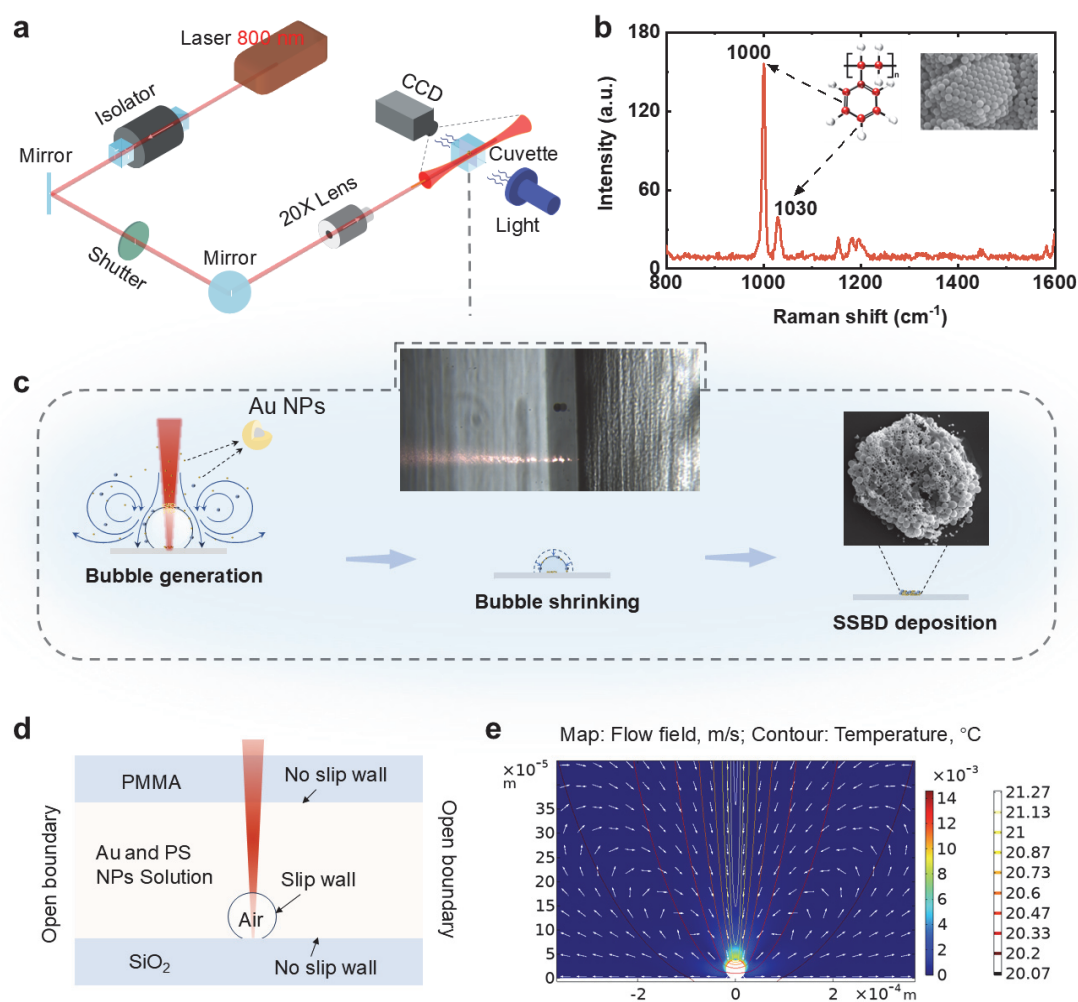

**Fig. 1 | Direct enrichment and detection of nanoplastics in water driven by the SSBD method.**
**a,** Optical setup for the SSBD process, mainly including 800 nm femtosecond pulsed laser, 20× objective lens, high-speed CCD camera, and PMMA cuvette. **b,** Characterization of pure PS NPs using SEM imaging (Supplementary Fig. 3) and Raman spectroscopy, showing diagnostic peaks at 1000 cm$^{-1}$ (benzene ring breathing) and 1030 cm$^{-1}$ (C–H in-plane bending). **c,** Schematics of the SSBD process: (1) optical pressure drives suspended SiO$_2$@Au NPs toward the glass slide to form the heat source; (2) a laser-generated thermal bubble induces Marangoni flow that transports NPs to the bubble surface; (3) concentrated NP deposition forms along the TPCL as the bubble shrinks and vanishes. Right inset: SEM image showing high-density deposition of 500 nm PS NPs from the PS solution. **d,** Schematic of the computational model and the applied boundary conditions for the finite-element simulation. **e,** Simulated Marangoni flow induced by laser heating of the suspension. Flow directions are indicated by arrows. The surface bubble has a diameter of ~40 μm, and the laser power is ~700 mW.

## Observation and quantification of 500, 200, and 30 nm PS nanoparticles
In the samples prepared by the SSBD method, we clearly observed dozens of 500 nm PS NPs concentrated in the deposition spot from a well-dispersed $10^3$ ng/mL PS NP suspension (Fig. 2a). The PS particles appear as roughly spherical shape, surrounded by co-deposited $SiO_2$@Au NPs. The intimate contact between $SiO_2$@Au NPs and PS NPs creates plasmonic hot spots, facilitating SERS enhancement for nanoplastics detection. Meanwhile, Raman mapping of the deposited sample enables approximate localization of PS NPs within the deposition area (Fig. 2b). The Raman spectrum collected from a representative point (Fig. 2c) clearly exhibits the symmetric ring breathing mode of the benzene ring at ~1000 $cm^{-1}$, along with an in-plane C–H bending vibration at ~1030 $cm^{-1}$ [43] [44], which matches well with reported reference spectra of PS [33]. The corresponding energy-dispersive X-ray (EDX) elemental mapping (Fig. 2d) further confirms the presence of PS NPs in SSBD deposition (Fig. 2a). Carbon signals overlap with the locations of the PS particles, while adjacent regions are enriched in gold from the $SiO_2$@Au NPs. Three representative points were selected for spectral comparison: one at a PS NP-rich area ($Y_1$), one at an $SiO_2$@Au NP-rich area ($Y_2$), and one at glass substrate ($Y_3$) (Fig. 2e). The spectra distinguish the three, with the PS sites showing a strong carbon peak at 0.277 keV, while oxygen (0.525 keV), silicon (1.73 keV), and gold (2.15 keV) peaks dominate in the $SiO_2$@Au NPs and glass substrate regions. The silicon peak also includes contributions from both the $SiO_2$@Au NPs and glass substrate.

Beyond qualitative identification, we performed quantitative analysis of 500 nm PS NPs with concentrations of $10^3$ ng/mL, $10^2$ ng/mL, 10 ng/mL, and 1 ng/mL (Fig. 2f). As expected, the number of deposited PS NPs decreased with concentration, showing an approximately linear relationship with the suspension concentration. The detection rate remained at 100% for concentrations $\geq 10^2$ ng/mL, meaning PS can be detected in every spot deposited. Here, the detection rate is defined as the ratio of spots in which PS NPs are successfully detected to the total number of analyzed spots (~8 per sample). It dropped to ~20% at 10 ng/mL, reflecting partial capture. No PS NPs were detectable at 1 ng/mL, establishing the LOD at 10 ng/mL. These results demonstrate that SSBD not only enables direct visualization and quantitative evaluation of nanoplastics but also reveals concentration-dependent capture efficiency. Specifically, at $10^3$ ng/mL, tens of 500 nm PS NPs were observed in each deposition spot, while several particles were detected at $10^2$ ng/mL. At 10 ng/mL, only one PS particle was occasionally captured in a SSBD spot, with some spots showing no PS NPs (Fig. 2a and Supplementary Fig. 4).

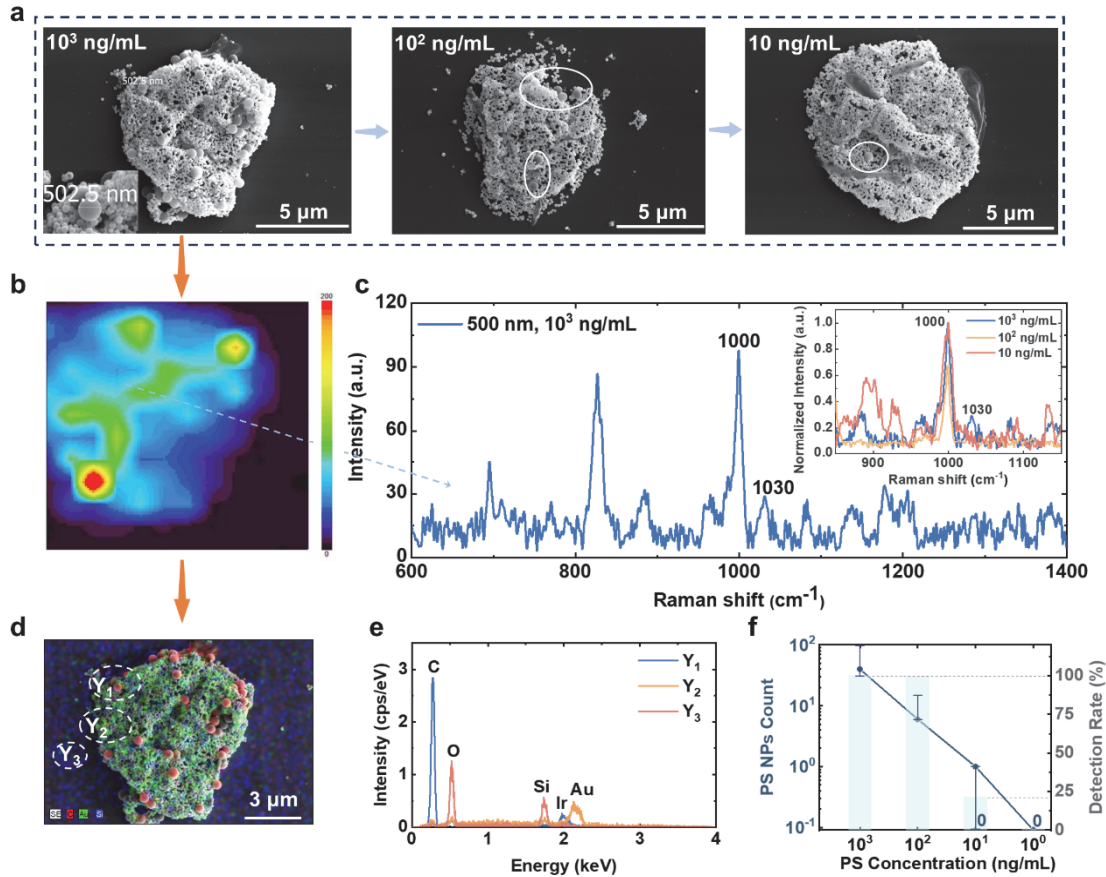

**Fig. 2 | Characterization and quantification of 500 nm PS nanoparticles at different concentrations. a,** SEM images of 500 nm PS-SiO$_2$@Au NPs deposition at different concentrations ($10^3$ ng/mL, $10^2$ ng/mL, and 10 ng/mL). **b,** Raman map at 1000 cm$^{-1}$ overlaid with the SEM image of $10^3$ ng/mL 500 nm PS-SiO$_2$@Au NPs deposition. **c,** SERS spectrum obtained from $10^3$ ng/mL 500 nm PS-SiO$_2$@Au NPs deposition. Inset: SERS normalized spectra obtained from $10^3$ ng/mL, $10^2$ ng/mL, 10 ng/mL cases. **d,** EDX elemental map overlaid with the SEM image of $10^3$ ng/mL 500 nm PS-SiO$_2$@Au NPs deposition. **e,** EDX spectra collected at three points, corresponding to carbon-, Si-, O-, and Au-rich regions, respectively. **f,** SSBD enrichment capacity at different concentrations, presented in terms of nanoparticle number in single deposition spot and detection rate across approximately eight deposition spots in one experiment.

To examine how particle size and number density influence the SSBD–SERS detection limit under the same mass concentration, we employed 200 nm PS NPs, which provided more than 15 times higher number density than the 500 nm PS NPs. As shown in Fig. 3a, these smaller PS particles were effectively concentrated within the deposition spot, resulting in a higher particle density in the enriched region. Figures 3a,b,c present SEM images of SSBD spots containing 200 nm PS NPs at different concentrations ($10^3$ ng/mL, 10 ng/mL, and $10^{-1}$ ng/mL, respectively). With decreasing concentration, the number of captured PS NPs decreases. Figure 3d also shows representative spectra exhibiting characteristic PS vibrational peaks at 1000 cm$^{-1}$ and 1030 cm$^{-1}$ for the $10^3$ ng/mL, 10 ng/mL, and $10^{-1}$ ng/mL 200 nm PS cases, consistent with the 500 nm PS cases in Fig. 2c. We further quantified a concentration series ($10^3$ ng/mL to $10^{-1}$ ng/mL) in Fig. 3e. At $10^3$ ng/mL, the deposition spot contained more 200 nm PS NPs than for the 500 nm PS NPs, which can be attributed to the higher particle number per volume for the same mass concentration. As the concentration decreased, the number of captured PS NPs remained sufficient for reliable Raman identification down to $10^{-1}$ ng/mL. Thus, for 200 nm PS NPs, the LOD of the SSBD-SERS improved to $10^{-1}$ ng/mL.

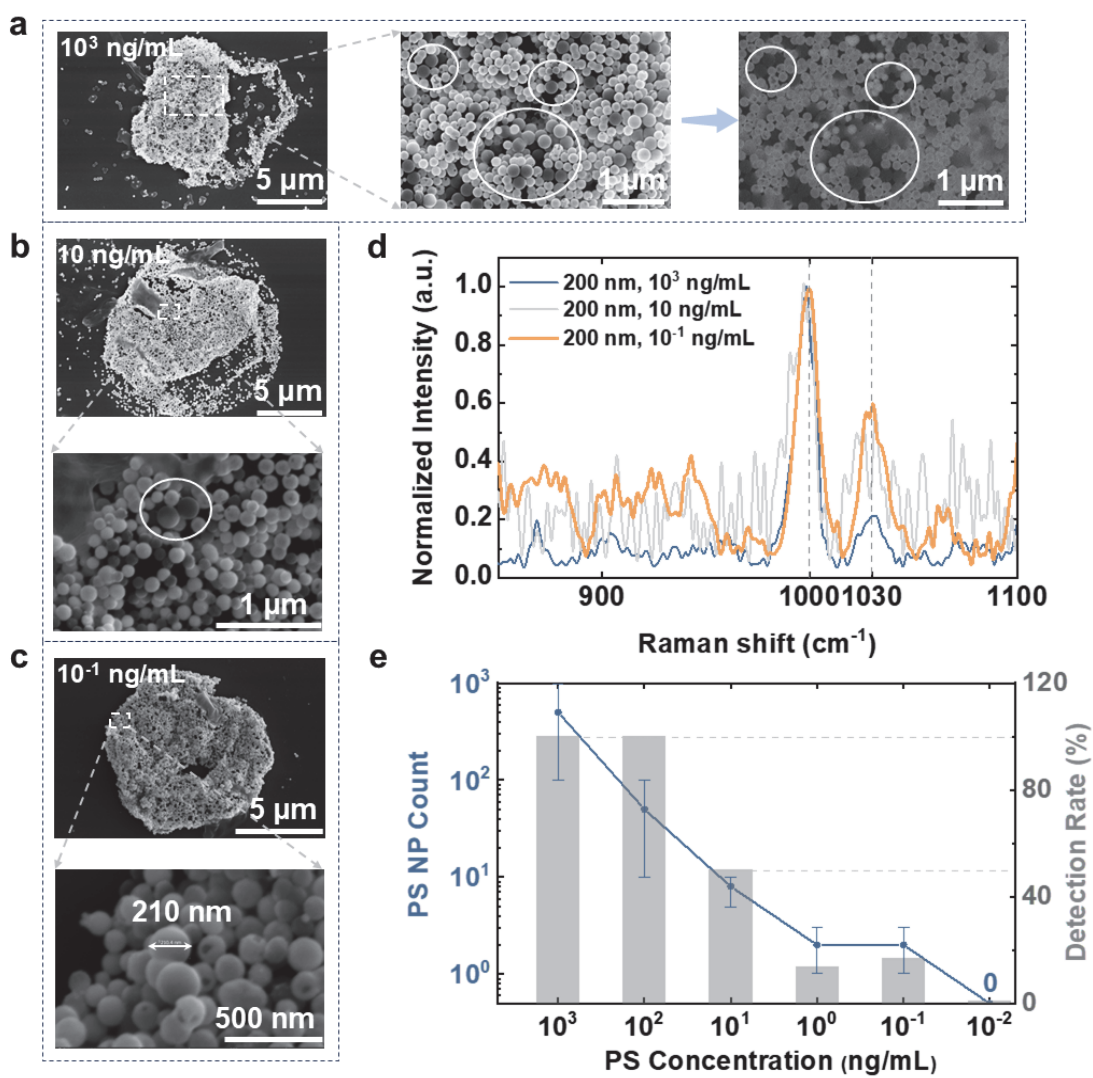

**Fig. 3 | Characterization and quantification of 200 nm PS nanoparticles at different concentrations. a,b,c,** SEM images of SSBD-prepared deposition spots of 200 nm PS NPs at different concentrations: (**a**) $10^3$ ng/mL, (**b**) 10 ng/mL, and (**c**) $10^{-1}$ ng/mL. With decreasing concentration, the number of captured particles decreases while maintaining the characteristic spherical morphology. **d,** SERS spectrum obtained from $10^3$ ng/mL, 10 ng/mL and $10^{-1}$ ng/mL 200 nm PS-SiO$_2$@Au NPs deposition. **e,** SSBD-enhanced detection sensitivity for 200 nm PS NPs as the concentration decreases, showing the number of PS NPs in a single deposition spot and the detection rate across multiple deposition spots within a single experiment.

Figure 4 presents the SERS spectra of 30 nm PS NPs at different concentrations, including $10^3$ ng/mL, 1 ng/mL, and $10^{-3}$ ng/mL, along with a systematic comparison of the characterization sensitivity for $10^3$ ng/mL 30 nm PS NPs using Raman SERS (Figs. 4a,b), SEM (Fig. 4c), and EDX (Figs. 4d,e,f). In Fig. 4a, the Raman map overlaid on the SEM image (Fig. 4c) reveals a spatial correlation between nanoparticle deposition and the Raman signal at the $10^3$ ng/mL case, supporting the capability of Raman mapping to identify and localize nanoplastics on the deposition. More importantly, as shown in Fig. 4b and Supplementary Fig. 5, the SERS spectra exhibit characteristic PS peaks at ~1000 and ~1030 cm$^{-1}$, which remain distinguishable at concentrations as low as $10^{-3}$ ng/mL, highlighting the high sensitivity of the SSBD-SERS platform. Meanwhile, the SEM images in Fig. 4c provide morphological information on the deposition of 30 nm PS nanoparticles at $10^3$ ng/mL. Only at magnifications as high as 125,000× can the particle size be clearly resolved.

However, accurate determination of surface coverage and particle clustering from SEM alone remains unfeasible, with such high-magnification images providing only qualitative confirmation of particle presence. For larger PS NPs (e.g., 200 nm or 500 nm), SEM provides much more effective characterization and quantitative analysis. However, detecting and distinguishing 30 nm PS NPs from background features on the ~10 μm scale becomes challenging, particularly in low-concentration samples such as $10^{-3}$ ng/mL. The EDX results shown in Figs. 4d,e,f further highlight these limitations. Neither the overlaid elemental distribution (Fig. 4e) nor the elemental carbon map (Fig. 4f) clearly reveals PS NP enrichment, as the signals appear spatially diffuse due to the large interaction volume of the electron beam. Because EDX spectra are averaged over micrometer-scale regions, individual or small numbers of 30 nm nanoparticles cannot be resolved, limiting the capability of EDX for ultrasensitive nanoplastic detection. Collectively, these results demonstrate that although SEM and EDX provide complementary morphological and elemental information, their effectiveness declines for nanoparticles below ~30 nm. In contrast, SERS offers clear chemical identification and quantitative detection across a broad concentration range, thereby advancing the LOD of SSBD method down to $10^{-3}$ ng/mL. This comparison underscores SSBD-SERS as a technique well-suited for trace-level monitoring of nanoplastics.

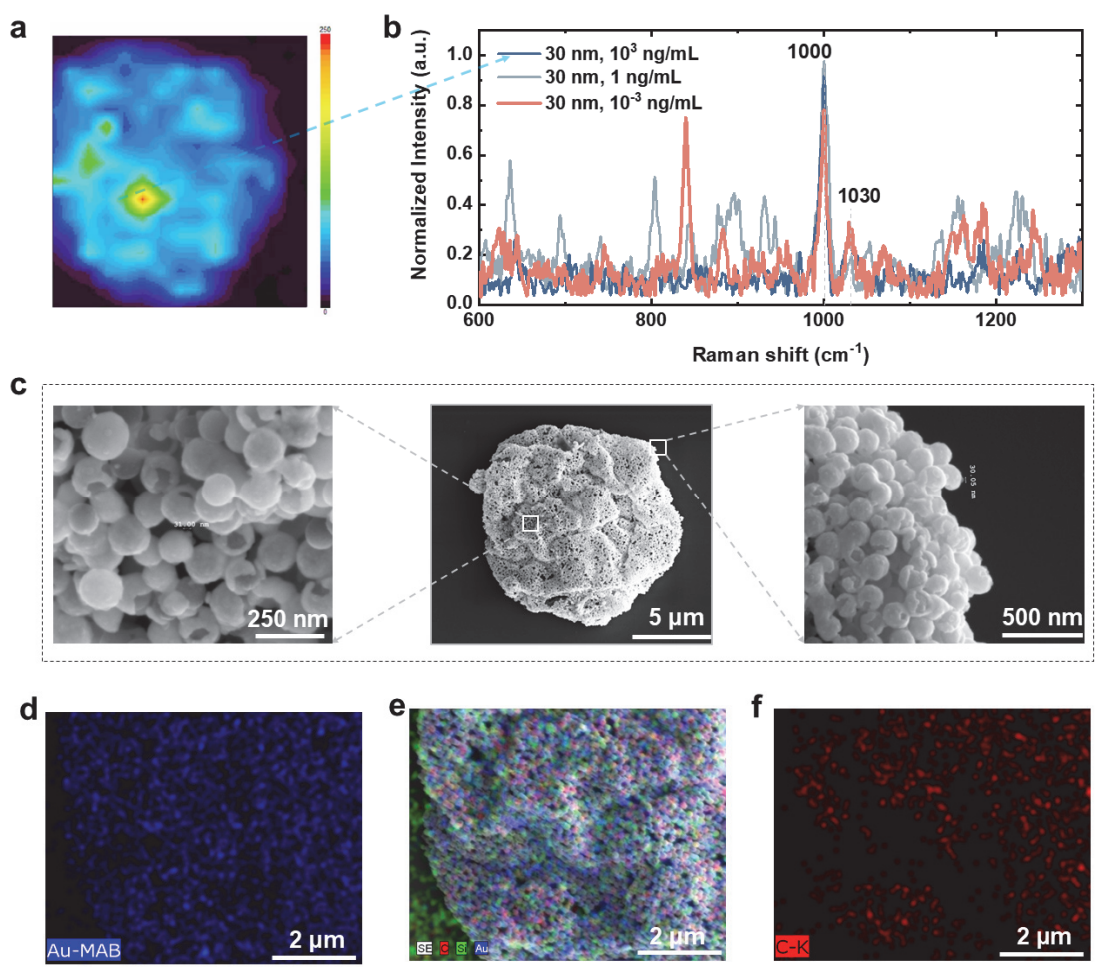

**Fig. 4 | Characterization of 30 nm PS nanoparticles at different concentrations. a,** Raman map at 1000 cm$^{-1}$ overlaid with the SEM image of 10$^3$ ng/mL 30 nm PS-SiO$_2$@Au NPs deposition. **b,** SERS spectra of 30 nm PS–SiO$_2$@Au NPs depositions at concentrations of 10$^3$ ng/mL, 1 ng/mL, and 10$^{-3}$ ng/mL. **c,** SEM images of 10$^3$ ng/mL 30 nm PS-SiO$_2$@Au NPs deposition, with corresponding high-magnification views. **d,e,f,** EDX elemental maps overlaid with the SEM image of a 10$^3$ ng/mL 30 nm PS–SiO$_2$@Au NP deposition. Individual color maps of Au (**d**) and C (**f**) elements correspond to panel (**e**).

## Comparison with existing methods

Combining the above observations across PS NPs of different sizes reveals a consistent trend: the LOD of the SSBD-SERS method decreases with decreasing PS particle diameter at a constant mass concentration. In our work, the SSBD-SERS method achieves LODs of 10 ng/mL for 500 nm PS NPs, 10$^{-1}$ ng/mL for 200 nm PS NPs, and 10$^{-3}$ ng/mL for 30 nm PS NPs, which are markedly lower than those reported for electro-photonic tweezers [45] and OM-SERS [33], as shown in Table 1. Specifically, a concentration of 10 ng/mL for 500 nm PS NPs corresponds to approximately 1.46 × 10$^5$ particles/mL, 10$^{-1}$ ng/mL for 200 nm PS NPs corresponds to about 2.27 × 10$^4$ particles/mL, and 10$^{-3}$ ng/mL for 30 nm PS NPs corresponds to approximately 6.74 × 10$^4$ particles/mL (Supplementary Fig. 6). These particle numbers were estimated by converting the total mass concentration to the number of particles per unit volume using the density of polystyrene ($\rho \approx 1.05$ g/cm$^3$) and the particle volume ($V = 4/3\pi r^3$).

These results suggest that the LOD of the SSBD-SERS method falls within the range of approximately $10^4$–$10^5$ particles/mL. The comparable particle counts across these size regimes indicate that the LOD of the SSBD–SERS method is primarily governed by the particle number density rather than by their mass concentration in suspension. A higher local particle number density enhances the probability of bubble-based enrichment and subsequent SERS signal generation. Consequently, this scaling relationship not only clarifies the size-dependent detection behavior and supports the generality of the SSBD-SERS mechanism across broad particle-size ranges, but also provides a potential basis for inferring the mass concentration of nanoplastics in unknown water samples.

**Table 1.** Reported LODs of the Raman-driven detection technologies for nanoplastics of different sizes.

| Approaches | Nanoplastics (size) | LODs (ng/mL) | References |
|---|---|---|---|
| Electro-photonic tweezers | PS (30 nm) | $1.0 \times 10^1$ | Yu et al., 2023 [45] |
| | PMMA (300 nm) | $1.0 \times 10^4$ | |
| OM-SERS | PS (30 nm) | $1.5 \times 10^{-1}$ | Shi et al., 2023 [33] |
| | PMMA (320 nm) | $8.0 \times 10^1$ | |
| | PET (750 nm) | $5.0 \times 10^1$ | |
| SSBD-SERS | PS (30 nm) | $1.0 \times 10^{-3}$ | The present work |
| | PS (200 nm) | $1.0 \times 10^{-1}$ | |
| | PS (500 nm) | $1.0 \times 10^1$ | |

**Detection of nanoplastics in drinking water**

Detecting nanoplastics in drinking water can be particularly challenging due to their low concentrations. While the exact concentrations vary depending on sampling location and analytical methodology, recent single-particle chemical imaging studies have estimated micro- and nanoplastic concentrations in bottled water to be around $1 \times 10^{-2}$ ng/mL, spanning a wide size range from micro- to nanoscale [25]. Notably, this concentration level is well within the detection capability of our SSBD–SERS platform, given its detection limit of $1 \times 10^{-3}$ ng/mL. Traditional optical or spectroscopic methods often fail to detect such low concentrations, making it important to establish techniques with lower detection thresholds. Our SSBD-SERS platform addresses this gap, enabling the capture and chemical identification of nanoplastics from complex aqueous matrices at concentrations below the detection limits of conventional approaches. To demonstrate its effectiveness, we investigated both bottled water and fountain water—two representative sources of daily human consumption.

Figure 5 consolidates the spectroscopic and microscopic evidence of micro- and nanoplastics identified in these samples. In the bottled water sample (Pure Life), the SERS spectrum from a hotspot (Fig. 5a, red) exhibits distinct vibrational features of aliphatic PA. These include C–C stretching modes between 1062–1130 cm$^{-1}$, an amide III band near 1300 cm$^{-1}$ (C–N stretching coupled with N–H bending), and a CH$_2$ scissoring vibration at 1446 cm$^{-1}$. SEM and EDX images (Figs. 5b,c) reveal nanoscale PA fragments with irregular morphologies deposited within the SSBD spot. Supporting evidence from Raman mapping at 1130 cm$^{-1}$, 1300 cm$^{-1}$, and 1446 cm$^{-1}$ (Supplementary Fig. 7) visualizes the spatial distribution of PA residues, with red-intensity regions indicating stronger SERS responses.

In contrast, fountain water collected from Fitzpatrick Hall at the University of Notre Dame displayed a distinct plastic signature. Besides PA, a distinct spectral signature consistent with PP was identified in the same water sample. As shown in Fig. 5d, the SERS spectrum of the captured particle (red) closely matches that of a PP reference material (blue). Prominent peaks at 810 and 842 cm$^{-1}$ correspond to CH$_2$ rocking vibrations typical of isotactic PP, while the band near 974 cm$^{-1}$ arises from symmetric CH$_3$ rocking. Additional modes between 1150–1168 cm$^{-1}$ (C–C stretching + in-plane CH bending) and around 1220 cm$^{-1}$ (CH wagging) further confirm the PP backbone structure. SEM and EDX images (Figs. 5e,f) display aggregated microscale PP particles with partially fused or irregular morphologies. Together, these results provide direct spectroscopic and morphological evidence of micro- and nanoplastics in drinking water, which demonstrate that the SSBD-SERS method can detect nanoplastics at concentrations below the thresholds of most traditional methods. This capability establishes the potential of SSBD as a platform for detecting trace-level nanoplastic contamination in real-world drinking water sources.

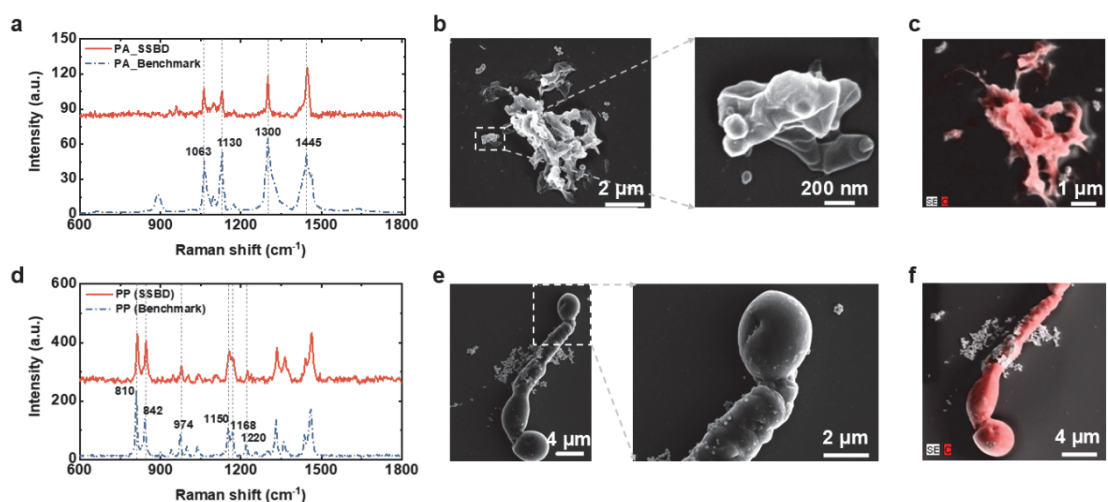

**Fig. 5 | Characterization and visualization of PA nanoplastics and PP microplastics in drinking water. a,** SERS spectra of bottled water sample compared with benchmark PA [46] [47], highlighting characteristic vibrational modes (C–C stretching, amide III, and CH$_2$ scissoring). **b,** SEM and **c,** EDX images of PA fragments collected on the SSBD substrate, revealing irregular nanoscale morphologies. **d,** SERS spectra of fountain water sample compared with the measured benchmark PP, with distinct aromatic ring vibration peaks confirming PP presence. **e,** SEM and **f,** EDX images of PP microplastics.

## Conclusions

In summary, we have demonstrated the effectiveness of the SSBD-SERS technique for enrichment and sensitive detection of nanoplastics in aqueous environments. By leveraging plasmonic bubble generation and Marangoni-driven thermofluidic flows, SSBD enables rapid and localized concentration of suspended NPs onto a microscale substrate area, facilitating their direct visualization and chemical identification via SERS. Our systematic quantification reveals that the detection limit of the SSBD-SERS technique depends on the particle number density, with smaller nanoplastics detectable at concentrations down to $10^{-3}$ ng/mL. Furthermore, polyamide and polypropylene micro/nanoplastics were successfully captured and analyzed in bottled and fountain water, respectively. The SSBD-SERS method overcomes the limitations of conventional ultrafiltration and optical manipulation approaches by combining efficient enrichment with minimal sample preparation while enabling chemical characterization. The ability to detect nanoplastics at concentrations as low as $10^{-3}$ ng/mL demonstrates SSBD's potential for environmental monitoring and human exposure risk assessment.

## Methods
### SSBD set-up and process
Testing samples for SSBD were prepared by mixing 500 μL of 150 nm $SiO_2$@Au nanoparticle suspension ($2 \times 10^8$ particles/mL) with 500 μL of diluted PS NPs solutions at various concentrations. The mixture was placed in a PMMA cuvette containing a clean glass slide. An 800-nm femtosecond pulsed laser (Spectra-Physics, Tsunami) was focused onto the glass slide through a 20× objective lens (NA 0.42). Laser irradiation was applied for ~20 s to induce photothermal bubble formation via plasmonic heating of the $SiO_2$@Au nanoparticles on the glass substrate. The 150 nm $SiO_2$@Au nanoparticles, with strong plasmonic resonance near 800 nm, facilitated rapid bubble nucleation at lower laser power (~700 mW). The surface bubble was allowed to grow to a diameter of ~40 μm before the laser was turned off, initiating bubble shrinkage. During shrinkage, suspended nanoplastics and $SiO_2$@Au NPs were concentrated at the three-phase contact line and deposited onto the substrate. The glass slide was then carefully removed and dried for 3 h prior to further analysis. Raman mapping was performed on the deposited sample to chemically identify nanoplastics, leveraging the SERS effect activated by the co-deposited plasmonic nanoparticles. Drinking water samples were prepared by mixing 500 μL of a 150 nm Au nanoparticle suspension ($2 \times 10^8$ particles/mL) with 500 μL of commercially bottled water (Pure Life) and fountain water. The mixture was processed using the same SSBD protocol described above.

### Electron microscopy settings
For SEM, a 5 nm iridium (Ir) conductive coating was applied onto SSBD deposition spots using an ACE600 Carbon & Sputter Coater. SEM imaging was performed with a Magellan 400 and Helios G4 UX at an accelerating voltage of 10 keV. Energy-dispersive X-ray spectroscopy (EDX) mappings were acquired at 4 keV with an approximate acquisition time of 50 s using a Bruker EDX system (Bruker Nano GmbH) integrated on the Magellan 400 SEM platform.

### Chemical characterization with raman measurements
Nanoplastics were chemically identified via their characteristic Raman vibrational bands using a confocal Raman microscope (NRS-5100, Jasco). The SERS effect arises from the localized surface plasmon resonance of the deposited $SiO_2$@Au NPs, producing intense nanoscale optical fields that enhance Raman signals of label-free nanoplastics. Raman mapping was performed with a 785 nm excitation laser and a 600 groove/mm grating at 1.0 μm intervals, unless otherwise noted. The laser power was set to ~0.2 mW (Supplementary Fig. 8), with spectral acquisition centered around 1050 cm$^{-1}$. For micro/nanoplastic detection in drinking water, Raman mapping was performed with a 532.1 nm excitation laser and a 1200 groove/mm. The laser power was set to ~1.2 mW, with spectral acquisition centered around 1091 cm$^{-1}$. Fluorescence background was automatically corrected for all measurements. A 100× objective lens was used for microfocusing and imaging. Nanoplastic locations were determined by overlaying Raman maps with optical images of the SSBD spot obtained under the same objective. SEM images of the full SSBD spot were aligned with the Raman mapping, enabling correlation between nanoplastic morphology and chemical signatures.

**Acknowledgments**

The authors would like to thank the support from the National Science Foundation (NSF) under grant number 2224307.

**Author contributions**

Y.L., S.M., and T.L. conceived the idea and designed the experiments. Y.L. R.Z., A.M. optimized and performed the SSBD process. Y.L. conducted Raman measurements. A.M. and R.Z. carried out the imaging process. S.M. and R.Z. conducted validation tests. Y.L., R.Z., and A.M. analyzed the experimental data. Y.L and A.M. conducted the theoretical calculation. The manuscript was initially composed by Y.L. R.Z., A.M., E.L., and S.M. All authors further contributed to the discussion of the experimental work and the final version of the manuscript.

**Competing interests**

The authors declare that they have no known competing financial interests or personal relationships that could have appeared to influence the work reported in this paper.

**Data and materials availability**

All data needed to evaluate the conclusions in the paper are present in the paper and/or the Supplementary Information.

**References**

1. Atugoda, T., Piyumali, H., Wijesekara, H., Sonne, C., Lam, S. S., Mahatantila, K., Vithanage, M. Nanoplastic occurrence, transformation and toxicity: a review. *Environmental Chemistry Letters* **21**, 363-381 (2023).
2. Chen, H., Xu, L., Yu, K., Wei, F., Zhang, M. Release of microplastics from disposable cups in daily use. *Science of The Total Environment* **854**, 158606 (2023).
3. Facciolà, A., Visalli, G., Pruiti Ciarello, M., Di Pietro, A. Newly emerging airborne pollutants: current knowledge of health impact of micro and nanoplastics. *International journal of environmental research and public health* **18**, 2997 (2021).
4. Zhao, J., Lan, R., Wang, Z., Su, W., Song, D., Xue, R., Liu, Z., Liu, X., Dai, Y., Yue, T. Microplastic fragmentation by rotifers in aquatic ecosystems contributes to global nanoplastic pollution. *Nat. Nanotechnol.* **19**, 406-414 (2024).
5. Lambert, S., Wagner, M. Characterisation of nanoplastics during the degradation of polystyrene. *Chemosphere* **145**, 265 (2016).
6. Pabortsava, K., Lampitt, R. S. High concentrations of plastic hidden beneath the surface of the

Atlantic Ocean. *Nat. Commun.* **11**, 4073 (2020).

7. Simpson, K., Martin, L., O'Leary, S. L., Watt, J., Moon, S., Luo, T., Xu, W. Environmental protein corona on nanoplastics altered the responses of skin keratinocytes and fibroblast cells to the particles. *Journal of Hazardous Materials* 138722 (2025).

8. Yi, J., Ma, Y., Ruan, J., You, S., Ma, J., Yu, H., Zhao, J., Zhang, K., Yang, Q., Jin, L. The invisible Threat: Assessing the reproductive and transgenerational impacts of micro-and nanoplastics on fish. *Environment international* **183**, 108432 (2024).

9. Dasmahapatra, A. K., Chatterjee, J., Tchounwou, P. B. A systematic review of the effects of nanoplastics on fish. *Frontiers in Toxicology* **7**, 1530209 (2025).

10. Thakur, R., Joshi, V., Sahoo, G. C., Jindal, N., Tiwari, R. R., Rana, S. Review of mechanisms and impacts of nanoplastic toxicity in aquatic organisms and potential impacts on human health. *Toxicology Reports* 102013 (2025).

11. Sarasamma, S., Audira, G., Siregar, P., Malhotra, N., Lai, Y.-H., Liang, S.-T., Chen, J.-R., Chen, K. H.-C., Hsiao, C.-D. Nanoplastics cause neurobehavioral impairments, reproductive and oxidative damages, and biomarker responses in zebrafish: throwing up alarms of wide spread health risk of exposure. *International journal of molecular sciences* **21**, 1410 (2020).

12. Mattsson, K., Johnson, E. V., Malmendal, A., Linse, S., Hansson, L.-A., Cedervall, T. Brain damage and behavioural disorders in fish induced by plastic nanoparticles delivered through the food chain. *Sci. Rep.* **7**, 11452 (2017).

13. Campen, M., Nihart, A., Garcia, M., Liu, R., Olewine, M., Castillo, E., Bleske, B., Scott, J., Howard, T., Gonzalez-Estrella, J. Bioaccumulation of microplastics in decedent human brains assessed by pyrolysis gas chromatography-mass spectrometry. *Research Square* rs. 3. rs-4345687 (2024).

14. Kaushik, A., Singh, A., Gupta, V. K., Mishra, Y. K. Nano/micro-plastic, an invisible threat getting into the brain. *Chemosphere* **361**, 142380 (2024).

15. Sun, Z., Wu, B., Yi, J., Yu, H., He, J., Teng, F., Xi, T., Zhao, J., Ruan, J., Xu, P. Impacts of environmental concentrations of nanoplastics on zebrafish neurobehavior and reproductive toxicity. *Toxics* **12**, 617 (2024).

16. Jayavel, S., Govindaraju, B., Michael, J. R., Viswanathan, B. Impacts of micro and nanoplastics on human health. *Bulletin of the National Research Centre* **48**, 110 (2024).

17. Barbosa, F., Adeyemi, J. A., Bocato, M. Z., Comas, A., Campiglia, A. A critical viewpoint on current issues, limitations, and future research needs on micro-and nanoplastic studies: From the detection to the toxicological assessment. *Environmental Research* **182**, 109089 (2020).

18. Zhang, R., Martin, L., Mandal, A., Liu, Y., Xu, J., Lee, E., Moon, S., Xu, W., Luo, T. A review of advancements and challenges in nanoplastics detection. *Cell Rep. Phys. Sci.* (2026).

19. Li, Q., Lai, Y., Li, P., Liu, X., Yao, Z., Liu, J., Yu, S. Evaluating the occurrence of polystyrene nanoparticles in environmental waters by agglomeration with alkylated ferroferric oxide followed by micropore membrane filtration collection and Py-GC/MS analysis. *Environmental Science & Technology* **56**, 8255-8265 (2022).

20. Shi, C., Liu, Z., Yu, B., Zhang, Y., Yang, H., Han, Y., Wang, B., Liu, Z., Zhang, H. Emergence of nanoplastics in the aquatic environment and possible impacts on aquatic organisms. *Science of the Total Environment* **906**, 167404 (2024).

21. Seeley, M. E., Lynch, J. M. Previous successes and untapped potential of pyrolysis–GC/MS for the analysis of plastic pollution. *Analytical and bioanalytical chemistry* **415**, 2873-2890 (2023).

22. Pei, W., Hu, R., Liu, H., Wang, L., Lai, Y. Advanced Raman spectroscopy for nanoplastics analysis: Progress and perspective. *TrAC Trends in Analytical Chemistry* **166**, 117188 (2023).
23. Dai, H., Li, H., Qiu, W., Deng, S., Han, J., Aminabhavi, T. Nondestructive analysis of plastic debris from micro to nano sizes: A state-of-the-art review on Raman spectroscopy-based techniques. *TrAC Trends in Analytical Chemistry* **176**, 117750 (2024).
24. Chang, L., Bai, S., Wei, P., Gao, X., Dong, J., Zhou, B., Peng, C., Jia, J., Luan, T. Quantitative detecting low concentration polystyrene nanoplastics in aquatic environments via an Ag/Nb2CTx (MXene) SERS substrate. *Talanta* **273**, 125859 (2024).
25. Qian, N., Gao, X., Lang, X., Deng, H., Bratu, T. M., Chen, Q., Stapleton, P., Yan, B., Min, W. Rapid single-particle chemical imaging of nanoplastics by SRS microscopy. *Proceedings of the National Academy of Sciences* **121**, e2300582121 (2024).
26. Mogha, N. K., Shin, D. Nanoplastic detection with surface enhanced Raman spectroscopy: Present and future. *TrAC Trends in Analytical Chemistry* **158**, 116885 (2023).
27. Mikac, L., Rigó, I., Himics, L., Tolić, A., Ivanda, M., Veres, M. Surface-enhanced Raman spectroscopy for the detection of microplastics. *Appl. Surf. Sci.* **608**, 155239 (2023).
28. Luo, S., Zhang, J., de Mello, J. C. Detection of environmental nanoplastics via surface-enhanced Raman spectroscopy using high-density, ring-shaped nanogap arrays. *Frontiers in Bioengineering and Biotechnology* **11**, 1242797 (2023).
29. Kousheh, S., Lin, M. Recent advancements in SERS-based detection of micro-and nanoplastics in food and beverages: techniques, instruments, and machine learning integration. *Trends Food Sci. Tech.* 104940 (2025).
30. Lê, Q. T., Ly, N. H., Kim, M.-K., Lim, S. H., Son, S. J., Zoh, K.-D., Joo, S.-W. Nanostructured Raman substrates for the sensitive detection of submicrometer-sized plastic pollutants in water. *Journal of hazardous materials* **402**, 123499 (2021).
31. Xu, G., Cheng, H., Jones, R., Feng, Y., Gong, K., Li, K., Fang, X., Tahir, M. A., Valev, V. K., Zhang, L. Surface-enhanced Raman spectroscopy facilitates the detection of microplastics< 1 μm in the environment. *Environmental science & technology* **54**, 15594-15603 (2020).
32. Ruan, X., Xie, L., Liu, J., Ge, Q., Liu, Y., Li, K., You, W., Huang, T., Zhang, L. Rapid detection of nanoplastics down to 20 nm in water by surface-enhanced raman spectroscopy. *Journal of Hazardous Materials* **462**, 132702 (2024).
33. Shi, X., Mao, T., Huang, X., Shi, H., Jiang, K., Lan, R., Zhao, H., Ma, J., Zhao, J., Xing, B. Capturing, enriching and detecting nanoplastics in water based on optical manipulation, surface-enhanced Raman scattering and microfluidics. *Nature Water* 1-12 (2025).
34. Moon, S., Martin, L. M., Kim, S., Zhang, Q., Zhang, R., Xu, W., Luo, T. Direct observation and identification of nanoplastics in ocean water. *Sci. Adv.* **10**, eadh1675 (2024).
35. Moon, S., Zhang, Q., Huang, D., Senapati, S., Chang, H. C., Lee, E., Luo, T. Biocompatible direct deposition of functionalized nanoparticles using shrinking surface plasmonic bubble. *Advanced Materials Interfaces* **7**, 2000597 (2020).
36. Zhang, Q., Neal, R. D., Huang, D., Neretina, S., Lee, E., Luo, T. Surface bubble growth in plasmonic nanoparticle suspension. *ACS Appl. Mater. Interfaces* **12**, 26680-26687 (2020).
37. Mandal, A., Zhang, Q., Zhang, R., Moon, S., Lee, E., Luo, T. Laser-Induced Trapping of Microbubbles within the Bulk Solution. *Langmuir* **41**, 19437-19443 (2025).
38. Lee, E., Huang, D., Luo, T. Ballistic supercavitating nanoparticles driven by single Gaussian beam optical pushing and pulling forces. *Nat. Commun.* **11**, 2404 (2020).

39. Lin, L., Peng, X., Mao, Z., Li, W., Yogeesh, M. N., Rajeeva, B. B., Perillo, E. P., Dunn, A. K., Akinwande, D., Zheng, Y. Bubble-pen lithography. *Nano Lett.* **16**, 701-708 (2016).
40. Fujii, S., Kanaizuka, K., Toyabe, S., Kobayashi, K., Muneyuki, E., Haga, M.-a. Fabrication and placement of a ring structure of nanoparticles by a laser-induced micronanobubble on a gold surface. *Langmuir* **27**, 8605-8610 (2011).
41. Zhang, Q., Li, R., Lee, E., Luo, T. Optically driven gold nanoparticles seed surface bubble nucleation in plasmonic suspension. *Nano Lett.* **21**, 5485-5492 (2021).
42. Moon, S., Zhang, Q., Xu, Z., Huang, D., Kim, S., Schiffbauer, J., Lee, E., Luo, T. Plasmonic nanobubbles–a perspective. *The Journal of Physical Chemistry C* **125**, 25357-25368 (2021).
43. Mayorga, C., Athalye, S. M., Boodaghidizaji, M., Sarathy, N., Hosseini, M., Ardekani, A., Verma, M. S. Limit of detection of Raman spectroscopy using polystyrene particles from 25 to 1000 nm in aqueous suspensions. *Analytical Chemistry* **97**, 8908-8914 (2025).
44. Nava, V., Frezzotti, M. L., Leoni, B. Raman spectroscopy for the analysis of microplastics in aquatic systems. *Applied Spectroscopy* **75**, 1341-1357 (2021).
45. Yu, E.-S., Jeong, E. T., Lee, S., Kim, I. S., Chung, S., Han, S., Choi, I., Ryu, Y.-S. Real-time underwater nanoplastic detection beyond the diffusion limit and low Raman scattering cross-section via electro-photonic tweezers. *ACS Nano* **17**, 2114-2123 (2022).
46. Uematsu, H., Kawasaki, T., Koizumi, K., Yamaguchi, A., Sugihara, S., Yamane, M., Kawabe, K., Ozaki, Y., Tanoue, S. Relationship between crystalline structure of polyamide 6 within carbon fibers and their mechanical properties studied using Micro-Raman spectroscopy. *Polymer* **223**, 123711 (2021).
47. Puchowicz, D., Cieslak, M., Puchowicz, D., Cieslak, M. Raman spectroscopy in the analysis of textile structures. *Recent developments in atomic force microscopy and Raman spectroscopy for materials characterization* 1-21 (2022).